\documentclass[11pt]{article}
\usepackage{amsmath}
\usepackage{graphicx}
\usepackage{enumerate}
\usepackage{etoolbox}
\usepackage{amsthm}
\usepackage{amssymb}
\usepackage{fullpage}
\usepackage{color}
\usepackage{mathtools}
\usepackage[utf8]{inputenc}
\usepackage[T1]{fontenc}
\usepackage{times}
\usepackage{caption}
\usepackage{subcaption}
\usepackage{abstract}
\usepackage{gensymb}
\usepackage{graphicx}
\usepackage{float}
\restylefloat{table}

\begin{document}
\renewcommand{\thefootnote}{\fnsymbol{footnote}}
\title{On closest isotropic tensors and their norms}
\author{
Tomasz Danek\footnote{Department of Geoinformatics and Applied Computer Science, AGH-University of Science and Technology, Krak\'ow, Poland, 
Email:~{\tt tdanek@agh.edu.pl}},
Andrea Noseworthy\footnote{Department of Earth Sciences,
Memorial University of Newfoundland, St.~John's, Newfoundland, Canada,\newline
Email:~{\tt anoseworthy@mun.ca}},
Michael A. Slawinski\footnote{Department of Earth Sciences,
Memorial University of Newfoundland, St.~John's, Newfoundland, Canada,\newline
Email:~{\tt mslawins@mac.ca}}
}
\date{\today}
\maketitle
\renewcommand{\thefootnote}{\arabic{footnote}}
\setcounter{footnote}{0}
\begin{abstract}
An anisotropic elasticity tensor can be approximated by the closest tensor belonging to a higher symmetry class.
The closeness of tensors depends on the choice of a criterion.
We compare the closest isotropic tensors obtained using four approaches: the Frobenius 36-component norm, the Frobenius 21-component norm, the operator norm and the $L_2$ slowness-curve fit.
We find that the isotropic tensors are similar to each other within the range of expected measurement errors.
\end{abstract}
\section{Introduction }
For an elasticity tensor obtained from empirical information, the resulting symmetry class is explicitly the property of a Hookean solid represented by that tensor, where this solid is a mathematical analogy of the physical material in question.
The inference of properties of that material requires further interpretation.
Among these properties there are symmetries of such a material, hence, it is useful to examine symmetries of its models.
In particular, it is useful to compute an isotropic counterpart of the obtained tensor.
The decision then lies in choosing an appropriate norm to compute the counterpart, hence the crux of this paper.
We compare isotropic counterparts according to the Frobenius-36 norm, Frobenius-21 norm, which we refer to as $F_{36}$ and $F_{21}$\,, respectively, as well as according to the operator norm and the $L_2$ slowness-curve fit, which we refer to as $\lambda$ and $L_2$\,, respectively.
\section{Elasticity tensors}
A Hookean solid, $c_{ijk\ell}$, is a mathematical object defined by Hooke's Law,\begin{equation}
\label{eq:Hooke}
\sigma_{ij}=\sum_{k=1}^{3}\sum_{\ell=1}^{3}c_{ijk\ell}\varepsilon_{k\ell}\,,\qquad i,j=1,2,3,
\end{equation}
where $\sigma_{ij}$\,, $\varepsilon_{k\ell}$ and $c_{ijk\ell}$ are the stress, strain and elasticity tensors, respectively.
The components of the elasticity tensor can be written---in Kelvin's, as opposed to Voigt's, notation---as a symmetric second-rank tensor in~${\mathbb R}^6$\,,
\begin{equation}
\label{eq:Elas}
C=
\left[\begin{matrix}
c_{1111}& c_{1122} & c_{1133} & \sqrt{2}c_{1123} & \sqrt{2}c_{1113} & \sqrt{2}c_{1112}\\
c_{1122}& c_{2222} & c_{2233} & \sqrt{2}c_{2223} & \sqrt{2}c_{2213} & \sqrt{2}c_{2212}\\
c_{1133}& c_{2233} & c_{3333} & \sqrt{2}c_{3323} & \sqrt{2}c_{3313} & \sqrt{2}c_{3312}\\
\sqrt{2}c_{1123}& \sqrt{2}c_{2223} & \sqrt{2}c_{3323} & 2c_{2323} & 2c_{2313} & 2c_{2312}\\
\sqrt{2}c_{1113}& \sqrt{2}c_{2213} & \sqrt{2}c_{3313} & 2c_{2313} & 2c_{1313} & 2c_{1312}\\
\sqrt{2}c_{1112}& \sqrt{2}c_{2212} & \sqrt{2}c_{3312} & 2c_{2312} & 2c_{1312} & 2c_{1212}\\
\end{matrix}\right]\,.
\end{equation}
For transverse isotropy, the components of $C$ become 
\begin{equation}
\label{eq:Trans}
C^{\rm TI}=
\left[\begin{matrix}
c_{1111}& c_{1122} & c_{1133} & 0 & 0 & 0\\
c_{1122}& c_{2222} & c_{2233} & 0 & 0 & 0\\
c_{1133}& c_{2233} & c_{3333} & 0 & 0 & 0\\
0& 0 & 0 & 2c_{2323} & 0 & 0\\
0& 0 & 0 & 0 & 2c_{2323} & 0\\
0& 0 & 0 & 0 & 0 & c_{1111}-c_{1122}\\
\end{matrix}\right]\,.
\end{equation}
For isotropy, the components of $C$ become
\begin{equation}
\label{eq:Iso}
C^{\rm iso}=
\left[\begin{matrix}
c_{1111}& c_{1111}-2c_{2323} & c_{1111}-2c_{2323} & 0 & 0 & 0\\
c_{1111}-2c_{2323}& c_{1111} & c_{1111}-2c_{2323} & 0 & 0 & 0\\
c_{1111}-2c_{2323}& c_{1111}-2c_{2323} & c_{1111} & 0 & 0 & 0\\
0& 0 & 0 & 2c_{2323} & 0 & 0\\
0& 0 & 0 & 0 & 2c_{2323} & 0\\
0& 0 & 0 & 0 & 0 & 2c_{2323}\\
\end{matrix}\right]\,,
\end{equation}
and expression~(\ref{eq:Hooke}) can be written as
\begin{equation*}
\sigma_{ij}=c_{1111}\,\delta_{ij}\sum_{k=1}^{3}\varepsilon_{kk}+2\,c_{2323}\,\varepsilon_{ij}\,,\qquad i,j=1,2,3\,.
\end{equation*}

\section{Norms}
\label{norms}
To examine the closeness between elasticity tensors, as discussed by Bos and Slawinski (2013) and by Danek et al. (2013, 2015), we consider possible norms of tensor~(\ref{eq:Elas}).
\subsection{Frobenius norms}
The Frobenius norm treats a matrix in $\mathbb{R}^{n\times n}$ as a Euclidean vector in $\mathbb{R}^{n^2}$\,.
In the case of a symmetric $6\times 6$ matrix, where $C_{mn}=C_{nm}$\,, we can choose either 
\begin{equation*}
||C||_{F_{36}} = \sqrt{\sum_{m=1}^6\sum_{n=1}^6 C_{mn}^{\,2}}
\,,
\end{equation*}
which uses the thirty-six components, including their coefficients of $\sqrt{2}$ and $2$\,, or
\begin{equation*}
||C||_{F_{21}} = \sqrt{\sum_{m=1}^{6}\sum_{n=1}^{m}C_{mn}^{\,2}}\,,
\end{equation*}
which uses only the twenty-one independent components, including their coefficients of $\sqrt{2}$ and $2$\,.

\subsection{Operator norm}
The operator norm of a symmetric $6\times 6$ matrix is
\begin{equation*}
||C||_{\lambda}=\max{|\lambda_i|}\,,
\end{equation*}
where $\lambda_i\in\{\lambda_1,\ldots,\lambda_6\}$\,, is an eigenvalue of $C$\,.
As discussed by Bos and Slawinski~(2013), such a norm of $c_{ijk\ell}$ results from equation~(\ref{eq:Hooke}) if $\sigma_{ij}$ and $\varepsilon_{k\ell}$ are {\it a~priori} endowed with a Frobenius norm.
If $c_{ijk\ell}$ is {\it a~priori} endowed with a Frobenius norm, its origin in the $\sigma_{ij}$ and $\varepsilon_{k\ell}$ norm does not exhibit any standard form.  
\section{Slowness-curve {\boldmath $L_2$} fit}
\label{misfit}
In a manner similar to the $F_{36}$~norm, $F_{21}$~norm and operator norm, the slowness-curve $L_2$ fit is used to find an isotropic counterpart of an anisotropic Hookean solid.
However, in contrast to these norms, which rely on finding the smallest distance between tensors, it relies on finding the best fit of circles—according to a chosen criterion—to noncircular wavefronts.

In this approach, in a manner similar to the operator norm, we do not invoke explicit expressions for the components of the closest elasticity tensor but we examine the effect of these components on certain quantities.
For the operator norm, this quantity consists of eigenvalues; for the slowness-curve fit, this quantity consists of wavefront slownesses.

The direct results of the norms are the components of the corresponding isotropic tensors, and the wavefront-slowness circles are their consequences.
The direct result of the slowness-curve fit are slowness circles, and the components of the corresponding isotropic tensor are their consequence.

The best fit, in the $L_2$ sense, is the radius,~$r$\,, that minimizes
\begin{equation}
\label{eq:misfit}
S=\sum_{i=1}^{n} (s_i-r_i)^2\,,
\end{equation}
where $s_i$ are $n$ discretized values along the slowness curve, and $s_i-r_i$ is measured in the radial direction.
Hence, $r$~is the radius of the slowness circle; it corresponds to isotropy.
\section{Numerical results}
\subsection{Tensor {\boldmath $C$}}

In this section, we investigate isotropic counterparts for the three norms introduced in Section~\ref{norms}.
For that purpose, we use a transversely isotropic tensor derived from a generally anisotropic tensor obtained by Dewangan and Grechka (2003),
\begin{equation}
\label{Aniso}
C=
\left[\begin{matrix}
7.8195& 3.4495 & 2.5667 & \sqrt{2}(0.1374) & \sqrt{2}(0.0558) & \sqrt{2}(0.1239)\\
3.4495& 8.1284 & 2.3589 & \sqrt{2}(0.0812) & \sqrt{2}(0.0735) & \sqrt{2}(0.1692)\\
2.5667& 2.3589 & 7.0908 & \sqrt{2}(-0.0092) & \sqrt{2}(0.0286) & \sqrt{2}(0.1655)\\
\sqrt{2}(0.1374)& \sqrt{2}(0.0812) & \sqrt{2}(-0.0092) & 2(1.6636) & 2(-0.0787) & 2(0.1053)\\
\sqrt{2}(0.0558)& \sqrt{2}(0.0735) & \sqrt{2}(0.0286) & 2(-0.0787) & 2(2.0660) & 2(-0.1517)\\
\sqrt{2}(0.1239)& \sqrt{2}(0.1692) & \sqrt{2}(0.1655) & 2(0.1053) & 2(-0.1517) & 2(2.4270)\\
\end{matrix}\right]\,.
\end{equation}
Its components are density-scaled elasticity parameters.
\subsection{Tensor {\boldmath $C_a^{\rm TI}$} and its isotropic counterparts}
\label{sub:TensorC}
\subsubsection{Tensor {\boldmath $C_a^{\rm TI}$}}
Let us consider a transversely isotropic tensor (Danek et al. 2013), which is the closest---in the $F_{36}$~sense---counterpart of tensor~(\ref{Aniso}),
\begin{equation}
\label{eq:ca}
C_a^{\rm TI}=
\left[\begin{matrix}
8.0641 & 3.3720 & 2.4588 & 0 & 0 & 0\\
3.3720 & 8.0641 & 2.4588 & 0 & 0 & 0\\
2.4588 & 2.4588 & 7.0817 & 0 & 0 & 0\\
0 & 0 & 0 & 2(1.8625) & 0 & 0\\
0 & 0 & 0 & 0 & 2(1.8625) & 0\\
0 & 0 & 0 & 0 & 0 & 2(2.3460)\\
\end{matrix}\right]\,.
\end{equation}
Isotropic tensors discussed herein are counterparts of this tensor.
The slowness curves for tensor~(\ref{eq:ca}) and its isotropic counterpart circles discussed in Sections~\ref{sec:F36}, \ref{sec:F21} and \ref{operator}, below, are shown in Figure~\ref{fig:ca}; these counterparts nearly coincide with each other.
\subsubsection{{\boldmath $F_{36}$} norm}
\label{sec:F36}
Let us consider the Frobenius norm using the thirty-six components.
There are analytical formul{\ae} to calculate---from a generally anisotropic tensor---the two parameters of its closest isotropic tensor (Voigt,~1910).
From a transversely isotropic tensor, these parameters are
\begin{equation*}
c_{1111}^{{\rm iso}_{F_{36}}} = \frac{1}{15}(8c_{1111}^{\rm TI} + 4c_{1133}^{\rm TI} + 8c_{2323}^{\rm TI} + 3c_{3333}^{\rm TI})
\end{equation*}
and
\begin{equation*}
c_{2323}^{{\rm iso}_{F_{36}}} = \frac{1}{15}(c_{1111}^{\rm TI}-2c_{1133}^{\rm TI}+5c_{1212}^{\rm TI}+6c_{2323}^{\rm TI}+c_{3333}^{\rm TI})\,.
\end{equation*}
Hence, the closest isotropic counterpart of tensor~(\ref{eq:ca}) is
\begin{equation}
\label{eq:caIso36}
C_a^{{\rm iso}_{F_{36}}}=
\left[\begin{matrix}
7.3662 & 2.9484 & 2.9484 & 0 & 0 & 0\\
2.9484 & 7.3662 & 2.9484 & 0 & 0 & 0\\
2.9484 & 2.9484 & 7.3662 & 0 & 0 & 0\\
0 & 0 & 0 & 2(2.2089) & 0 & 0\\
0 & 0 & 0 & 0 & 2(2.2089) & 0\\
0 & 0 & 0 & 0 & 0 & 2(2.2089)\\
\end{matrix}\right]\,.
\end{equation}
\subsubsection{{\boldmath $F_{21}$} norm}
\label{sec:F21}
Let us consider the Frobenius norm using the twenty-one independent components.
The analytical formul{\ae} to calculate the two parameters of its closest isotropic tensor are (Slawinski,~2016)
\begin{equation*}
c_{1111}^{{\rm iso}_{F_{21}}}=\frac{1}{9}(-c_{1122}^{\rm TI}+2(3c_{2222}^{\rm TI}+c_{2233}^{\rm TI}+2c_{2323}^{\rm TI}+c_{3333}^{\rm TI}))
\end{equation*}
and
\begin{equation*}
c_{2323}^{{\rm iso}_{F_{21}}}=\frac{1}{18}(-5c_{1122}^{\rm TI}+6c_{2222}^{\rm TI}-2c_{2233}^{\rm TI}+8c_{2323}^{\rm TI}+c_{3333}^{\rm TI})\,.
\end{equation*}
Hence,
\begin{equation}
\label{eq:caIso21}
C_a^{{\rm iso}_{F_{21}}}=
\left[\begin{matrix}
7.4279 & 3.0716 & 3.0716 & 0 & 0 & 0\\
3.0716 & 7.4279 & 3.0716 & 0 & 0 & 0\\
3.0716 & 3.0716 & 7.4279 & 0 & 0 & 0\\
0 & 0 & 0 & 2(2.1781) & 0 & 0\\
0 & 0 & 0 & 0 & 2(2.1781) & 0\\
0 & 0 & 0 & 0 & 0 & 2(2.1781)\\
\end{matrix}\right]\,.
\end{equation}
\subsubsection{{\boldmath $\lambda$} norm}
\label{operator}
Unlike the Frobenius norms, the operator norm has no analytical formul{\ae} for $\rm c_{1111}^{\rm iso_{\lambda}}$ and $\rm c_{2323}^{\rm iso_{\lambda}}$\,.
They must be obtained numerically.
For tensor~(\ref{eq:ca}), we obtain
\begin{equation}
\label{eq:caIsoL2}
C_a^{{\rm iso}_{\lambda}}=
\left[\begin{matrix}
7.7562 & 3.0053 & 3.0053 & 0 & 0 & 0\\
3.0053 & 7.7562 & 3.0053 & 0 & 0 & 0\\
3.0053 & 3.0053 & 7.7562 & 0 & 0 & 0\\
0 & 0 & 0 & 2(2.3755) & 0 & 0\\
0 & 0 & 0 & 0 & 2(2.3755) & 0\\
0 & 0 & 0 & 0 & 0 & 2(2.3755)\\
\end{matrix}\right]\,.
\end{equation}

\begin{figure}[ht]
  \centering
  \begin{minipage}[t]{0.4\textwidth}
    \includegraphics[width=\textwidth]{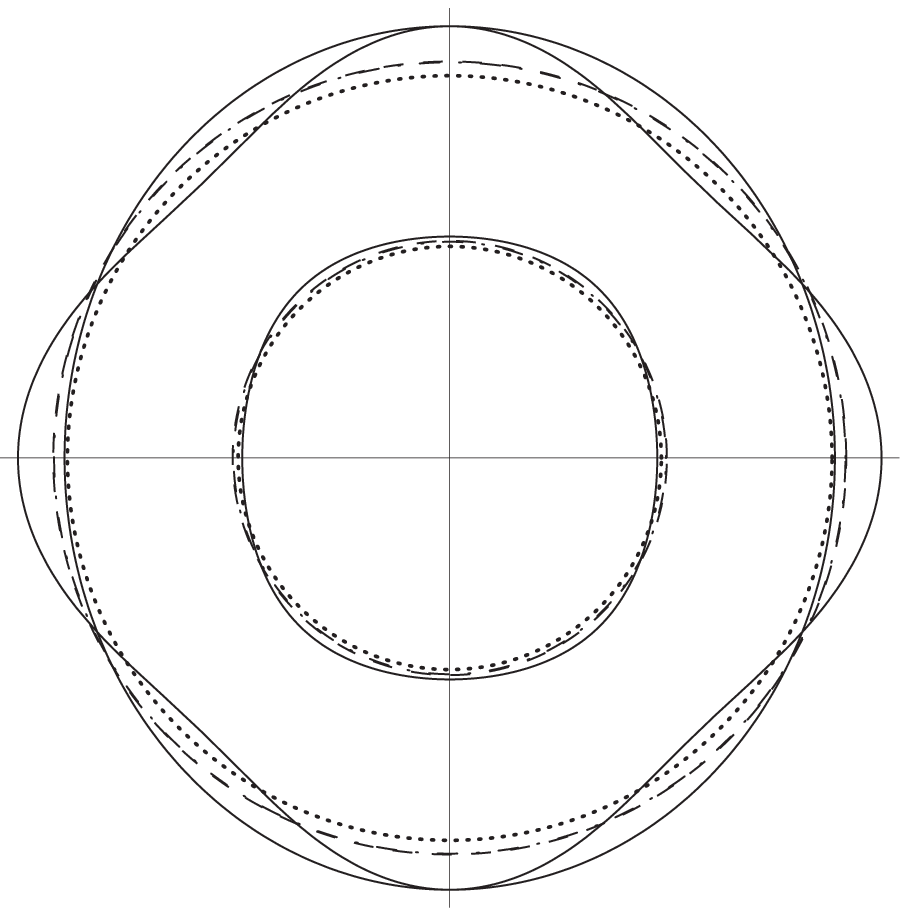}
    \caption{Slowness curves for tensor~(\ref{eq:ca}):
solid lines represent the $qP$\,, $qSV$ and $SH$ waves;
dashed lines represent the $P$ and $S$ waves according to $F_{36}$~norm;
dashed-dotted lines represent the $P$ and $S$ waves according to $F_{21}$~norm;
the results of these norms almost coincide;
dotted lines represent the $P$ and $S$ waves according to $\lambda$~norm.}
\label{fig:ca}
  \end{minipage}
  \hfill
  \begin{minipage}[t]{0.4\textwidth}
    \includegraphics[width=\textwidth]{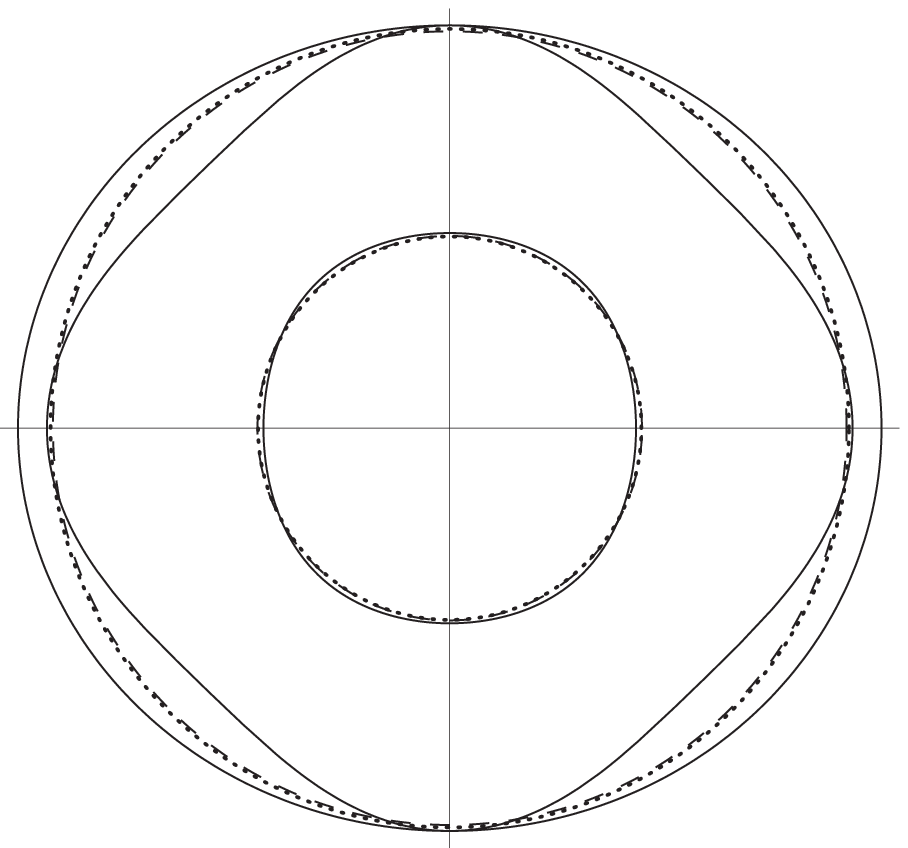}
    \caption{Slowness curves for tensor~(\ref{eq:cb}):
solid lines represent the $qP$\,, $qSV$ and $SH$ waves;
dashed lines represent the $P$ and $S$ waves according to $F_{36}$~norm;
dotted lines represent the $P$ and $S$ waves according to $F_{21}$~norm.}
\label{fig:cb}
  \end{minipage}
\end{figure}
\subsubsection{Distances among tensors}
To gain insight into different isotropic counterparts of tensor~(\ref{eq:ca}), we calculate the $F_{36}$ distance between tensors~(\ref{eq:caIso36}) and (\ref{eq:caIsoL2}), which is $0.8993$\,.
The $F_{36}$ distance between tensors~(\ref{eq:ca}) and (\ref{eq:caIso36}) is $1.8461$\,.
The $F_{36}$ distance between tensors~(\ref{eq:ca}) and (\ref{eq:caIsoL2}) is $2.0535$\,, where we note that tensor~(\ref{eq:caIsoL2}) is the closest isotropic tensor according to the operator---not the $F_{36}$---norm. 
Thus, in spite of similarities between the isotropic tensors, the distance between them is large in comparison to their distances to tensor~(\ref{eq:ca}).

This is an illustration of abstractness of the concept of distances in the space of elasticity tensors.
A concrete evaluation is provided by comparing the results obtained by minimizing these distances.
Such results are tensors~(\ref{eq:caIso36}), (\ref{eq:caIso21}), (\ref{eq:caIsoL2}), and their wavefront-slowness circles in Figure~\ref{fig:ca}.
This figure illustrates a similarity among these circles, which is a realm in which the isotropic tensors can be compared.
They can be compared within the slowness space.
\subsection{Comparison of norms}
\label{sub:Comparison}
Comparing tensors~(\ref{eq:caIso36}), (\ref{eq:caIso21}) and (\ref{eq:caIsoL2}), we see that the parameters of the closest isotropic tensor depend on the norm used.
Given two anisotropic tensors, we might be interested to know which of them is closer to isotropy.
For a given norm, a unique answer is obtained by a straightforward calculation.
For different norms, there is no unique answer: the sequence in closeness to isotropy can be reversed between two tensors.
\subsubsection{{\boldmath $F_{36}$} versus {\boldmath $F_{21}$}}
Using a numerical search, an elasticity tensor is generated that is further from isotropy than tensor~(\ref{eq:ca}) according to the $F_{36}$~norm, but closer to isotropy than tensor~(\ref{eq:ca}) according to the $F_{21}$~norm. The search results in
\begin{equation}
\label{eq:cb}
C_b^{\rm TI}=
\left[\begin{matrix}
7.3091 & 4.5882 & 2.9970 & 0 & 0 & 0\\
4.5882 & 7.3091 & 2.9970 & 0 & 0 & 0\\
2.9970 & 2.9970 & 6.6604 & 0 & 0 & 0\\
0 & 0 & 0 & 2(1.5631) & 0 & 0\\
0 & 0 & 0 & 0 & 2(1.5631) & 0\\
0 & 0 & 0 & 0 & 0 & 2(1.3605)\\
\end{matrix}\right]\,,
\end{equation}
with its corresponding isotropic counterparts,
\begin{equation}
\label{eq:cbIso36}
C_b^{{\rm iso}_{F_{36}}}=
\left[\begin{matrix}
6.8631 & 3.6422 & 3.6422 & 0 & 0 & 0\\
3.6422 & 6.8631 & 3.6422 & 0 & 0 & 0\\
3.6422 & 3.6422 & 6.8631 & 0 & 0 & 0\\
0 & 0 & 0 & 2(1.6104) & 0 & 0\\
0 & 0 & 0 & 0 & 2(1.6104) & 0\\
0 & 0 & 0 & 0 & 0 & 2(1.6104)\\
\end{matrix}\right]
\end{equation}
and
\begin{equation}
\label{eq:cbIso21}
C_b^{{\rm iso}_{F_{21}}}=
\left[\begin{matrix}
6.9014 & 3.7188 & 3.7188 & 0 & 0 & 0\\
3.7188 & 6.9014 & 3.7188 & 0 & 0 & 0\\
3.7188 & 3.7188 & 6.9014 & 0 & 0 & 0\\
0 & 0 & 0 & 2(1.5913) & 0 & 0\\
0 & 0 & 0 & 0 & 2(1.5913) & 0\\
0 & 0 & 0 & 0 & 0 & 2(1.5913)\\
\end{matrix}\right]\,,
\end{equation}
respectively.
The distances to isotropy for $C_a^{\rm TI}$ and $C_b^{\rm TI}$\,, using the $F_{36}$ and $F_{21}$~norms, are$$d_{a_{21}}=1.6372 > d_{b_{21}}=1.5517\,,$$$$d_{a_{36}}= 1.8460 < d_{b_{36}}=2.0400\,.$$
The slowness curves for tensor~(\ref{eq:cb}) and its isotropic counterparts are shown in Figure~\ref{fig:cb}.
\subsubsection{{\boldmath $F_{36}$} versus {\boldmath $\lambda$}}
The second comparison is between the $F_{36}$~norm and the $\lambda$~norm. We obtain 
\begin{equation}
\label{eq:cbb}
C_{bb}^{\rm TI}=
\left[\begin{matrix}
6.8639 & 3.3046 & 2.8770 & 0 & 0 & 0\\
3.3046 & 6.8639 & 2.8770 & 0 & 0 & 0\\
2.8770 & 2.8770 & 8.3825 & 0 & 0 & 0\\
0 & 0 & 0 & 2(2.7744) & 0 & 0\\
0 & 0 & 0 & 0 & 2(2.7744) & 0\\
0 & 0 & 0 & 0 & 0 & 2(1.7797)\\
\end{matrix}\right]\,,
\end{equation}
which is further from isotropy according to the $F_{36}$~norm and closer to isotropy according to the $\lambda$~norm.
Its isotropic counterparts in the sense of the $F_{36}$ and $\lambda$~norms are
\begin{equation}
\label{eq:cbbIso36}
C_{bb}^{{\rm iso}_{F_{36}}}=
\left[\begin{matrix}
7.5842 & 2.9125 & 2.9125 & 0 & 0 & 0\\
2.9125 & 7.5842 & 2.9125 & 0 & 0 & 0\\
2.9125 & 2.9125 & 7.5842 & 0 & 0 & 0\\
0 & 0 & 0 & 2(2.3358) & 0 & 0\\
0 & 0 & 0 & 0 & 2(2.3358) & 0\\
0 & 0 & 0 & 0 & 0 & 2(2.3358)\\
\end{matrix}\right]
\end{equation}
and
\begin{equation}
\label{eq:cbbIsoL2}
C_{bb}^{{\rm iso}_{\lambda}}=
\left[\begin{matrix}
7.4712 & 2.9171 & 2.9171 & 0 & 0 & 0\\
2.9171 & 7.4712 & 2.9171 & 0 & 0 & 0\\
2.9171 & 2.9171 & 7.4712 & 0 & 0 & 0\\
0 & 0 & 0 & 2(2.7704) & 0 & 0\\
0 & 0 & 0 & 0 & 2(2.7704) & 0\\
0 & 0 & 0 & 0 & 0 & 2(2.7704)\\
\end{matrix}\right]\,,
\end{equation}
respectively.
The distances to isotropy for $C_a^{\rm TI}$ and $C_{bb}^{\rm TI}$\,, using the $F_{36}$ and $\lambda$~norms, are $$d_{a_{36}}=1.8460 < d_{bb_{36}}=2.1825\,,$$$$d_{a_{\lambda}}= 1.0259 > d_{bb_{\lambda}}=0.9947\,.$$
The slowness curves for tensor~(\ref{eq:cbb}) and its isotropic counterparts are shown in Figure~\ref{fig:cbb}.
\subsubsection{{\boldmath $F_{21}$} versus {\boldmath $\lambda$}}
\begin{figure}[!tbp]
\centering
\begin{minipage}[t]{0.4\textwidth}
    \includegraphics[width=\textwidth]{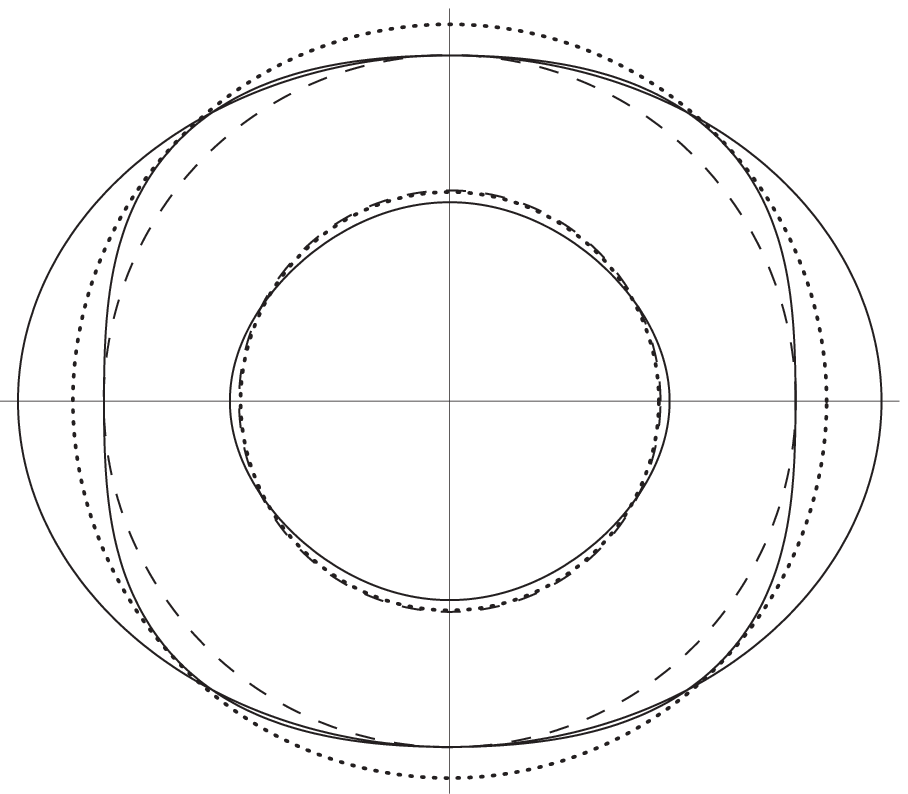}
    \caption{Slowness curves for tensor~(\ref{eq:cbb}):
solid lines represent the $qP$\,, $qSV$ and $SH$ waves;
dotted lines represent its $P$ and $S$ waves according to $F_{36}$~norm;
dashed lines represent its $P$ and $S$ waves according to $\lambda$~norm.}
\label{fig:cbb}
  \end{minipage}
\hfill
\begin{minipage}[t]{0.4\textwidth}
  \centering
     \includegraphics[width=\textwidth]{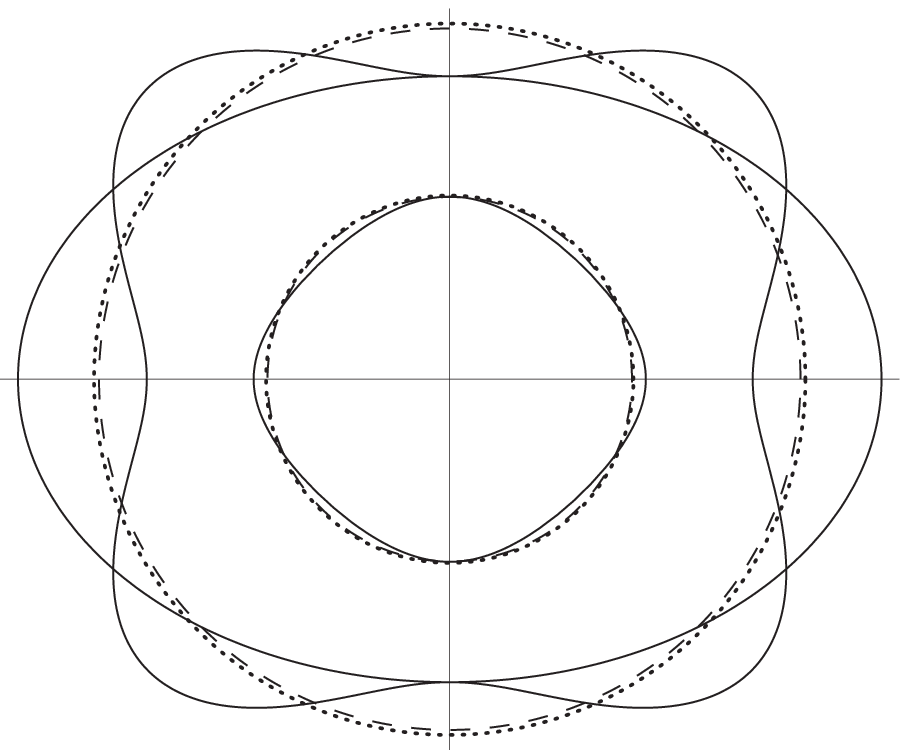}
    \caption{Slowness curves for tensor~(\ref{eq:cbbb}):
solid lines represent the $qP$\,, $qSV$ and $SH$ waves;
dotted lines represent its $P$ and $S$ waves according to $F_{21}$~norm;
dashed lines represent its $P$ and $S$ waves according to $\lambda$~norm.}
\label{fig:cbbb}
  \end{minipage}
\end{figure}

The third comparison is between the $F_{21}$~norm and the $\lambda$~norm.
The resulting tensor is
\begin{equation}
\label{eq:cbbb}
C_{bbb}^{\rm TI}=
\left[\begin{matrix}
4.5706 & 2.6852 & 2.9075 & 0 & 0 & 0\\
2.6852 & 4.5706 & 2.9075 & 0 & 0 & 0\\
2.9075 & 2.9075 & 5.2705 & 0 & 0 & 0\\
0 & 0 & 0 & 2(1.9145) & 0 & 0\\
0 & 0 & 0 & 0 & 2(1.9145) & 0\\
0 & 0 & 0 & 0 & 0 & 2(0.9427)\\
\end{matrix}\right]\,,
\end{equation}
with isotropic counterparts according to the $F_{21}$~norm and the $\lambda$~norm,
\begin{equation}
\label{eq:cbbbIso21}
C_{bbb}^{{\rm iso}_{F_{21}}}=
\left[\begin{matrix}
5.2074 & 2.4297 & 2.4297 & 0 & 0 & 0\\
2.4297 & 5.2074 & 2.4297 & 0 & 0 & 0\\
2.4297 & 2.4297 & 5.2074 & 0 & 0 & 0\\
0 & 0 & 0 & 2(1.3889) & 0 & 0\\
0 & 0 & 0 & 0 & 2(1.3889) & 0\\
0 & 0 & 0 & 0 & 0 & 2(1.3889)\\
\end{matrix}\right]
\end{equation}
and
\begin{equation}
\label{eq:cbbbIsoL2}
C_{bbb}^{{\rm iso}_{\lambda}}=
\left[\begin{matrix}
5.2926 & 2.4354 & 2.4354 & 0 & 0 & 0\\
2.4354 & 5.2926 & 2.4354 & 0 & 0 & 0\\
2.4354 & 2.4354 & 5.2926 & 0 & 0 & 0\\
0 & 0 & 0 & 2(1.4286) & 0 & 0\\
0 & 0 & 0 & 0 & 2(1.4286) & 0\\
0 & 0 & 0 & 0 & 0 & 2(1.4286)\\
\end{matrix}\right]\,,
\end{equation}
respectively. The distances to isotropy for both $C_a^{\rm TI}$ and $C_{bbb}^{\rm TI}$ using the $F_{21}$ and $\lambda$~norms are $$d_{a_{21}}=1.6372 < d_{{bbb}_{21}}=2.0842\,,$$$$d_{a_{\lambda}}= 1.0259 > d_{bbb_{\lambda}}=0.9719\,.$$
The slowness curves for tensor~(\ref{eq:cbbb}) and its isotropic counterparts are shown in Figure~\ref{fig:cbbb}.
\subsection{Slowness-curve fit}
\label{sub:Misfit}
Considering tensor~(\ref{eq:ca}) and applying a minimization for the $qP$ wave, using formula~(\ref{eq:misfit}), we find ${S=0.0886}$ with ${r=0.3770}$\,.
Following the same procedure for the  $qSV$ and $SH$ waves, we find ${S=0.2973}$\,, with ${r=0.6832}$\,, and ${S=0.2169}$\,, with ${r=0.6831}$\,, respectively.
Combining these results, we obtain ${S=0.6029}$\,, with ${r_P=0.3770}$ and ${r_S=0.6831}$\,, which are the slownesses of the $P$ and $S$ waves, respectively.
Note that---since the slowness curves of the $qP$ waves are detached from the curves for the $qSV$ and $SH$ waves---the value of $r$ for the $P$ waves does not change by combining the results.

\begin{figure}[!htb]
  \centering
    \includegraphics[width=6cm]{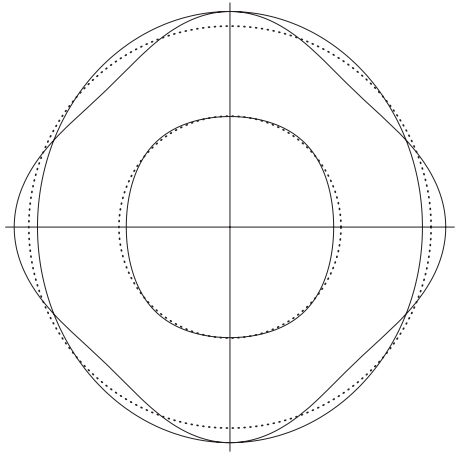}
    \caption{Slowness curves for tensor~(\ref{eq:aMisfit}):
solid lines represent the $qP$\,, $qSV$ and $SH$ waves;
dotted lines represent its $P$ and $S$ waves according to the slowness-curve $L_2$ fit.}
\label{fig:misfit}
  \end{figure}

Since $v_P=\sqrt{c_{1111}}$ and $v_S=\sqrt{c_{2323}}$ are the $P$-wave and $S$-wave speeds, respectively, it follows that $c_{1111}=1/r_P^2$ and $c_{2323}=1/r_S^2$\,.
Hence, we obtain
\begin{equation}
\label{eq:aMisfit}
C_a^{{\rm iso}_{L_2}}=
\left[\begin{matrix}
7.0341 & 2.7485 & 2.7485 & 0 & 0 & 0\\
2.7485 & 7.0341 & 2.7485 & 0 & 0 & 0\\
2.7485 & 2.7485 & 7.0341 & 0 & 0 & 0\\
0 & 0 & 0 & 2(2.1428) & 0 & 0\\
0 & 0 & 0 & 0 & 2(2.1428) & 0\\
0 & 0 & 0 & 0 & 0 & 2(2.1428)\\
\end{matrix}\right]\,.
\end{equation}
The slowness curves for tensor~(\ref{eq:aMisfit}) and its isotropic counterparts are shown in Figure~\ref{fig:misfit}.

\subsection{Thomsen parameters}
\label{sub:Thomsen}
Tensors~(\ref{eq:ca}), (\ref{eq:cb}), (\ref{eq:cbb}) and (\ref{eq:cbbb}) exhibit the strength of anisotropy that is consistent with cases of interest to geophysicists.
To show this consistency, we calculate the Thomsen (1986) parameters using the following formul{\ae},
\begin{align*}
\alpha &= \sqrt{c_{3333}^{\rm TI}}\,,\\
\beta &= \sqrt{c_{2323}^{\rm TI}}\,,\\
\gamma &= \frac{c_{1212}^{\rm TI}-c_{2323}^{\rm TI}}{2c_{2323}^{\rm TI}}\,,\\
\delta &= \frac{(c_{1133}^{\rm TI}+c_{2323}^{\rm TI})^2-(c_{3333}^{\rm TI}-c_{2323}^{\rm TI})^2}{2c_{3333}^{\rm TI}(c_{3333}^{\rm TI}-c_{2323}^{\rm TI})}\,,\\
\epsilon &= \frac{c_{1111}^{\rm TI}-c_{3333}^{\rm TI}}{2c_{3333}^{\rm TI}}\,.
\end{align*}
The values of these parameters for tensors~(\ref{eq:ca}), (\ref{eq:cb}), (\ref{eq:cbb}) and (\ref{eq:cbbb}) are shown in Table~\ref{table:Thomsen}. Comparing results of this table to data of Auld~(1973) and Thomsen~(1986), we see that these tensors can represent common geological materials.

\begin{table}[H]
\caption{Thomsen parameters for tensors~(\ref{eq:ca}), (\ref{eq:cb}), (\ref{eq:cbb}) and (\ref{eq:cbbb})}
\label{table:Thomsen}
\begin{center}
 \begin{tabular}{||c|c|c|c|c|c||}
 \hline
 Tensor & $\rm\alpha$ & $\rm\beta$ & $\rm\gamma$ & $\rm\delta$ & $\rm\varepsilon$ \\ [0.5ex] 
 \hline\hline
 $C_a^{\rm TI}$ & 2.6612 & 1.2986 & 0.1956 & -0.1561 & 0.0694\\ 
 \hline
 $C_b^{\rm TI}$ & 2.5808 & 1.2503 & -0.6483 & -0.0764 & 0.0487\\
 \hline
 $C_{bb}^{\rm TI}$ & 2.2958 & 1.3837 & -0.2538 & 0.3389 & -0.6640\\
 \hline
 $C_{bbb}^{\rm TI}$ & 2.8953 & 1.6657 & -0.1793 & 0.0052 & -0.0906\\
 \hline 
 \hline
\end{tabular}
\end{center}
\end{table}
\subsection{Error propagation}
\label{sub:Error}
\begin{figure}[!htb]
\centering
  \begin{minipage}[b]{0.3\textwidth}
    \includegraphics[width=5cm]{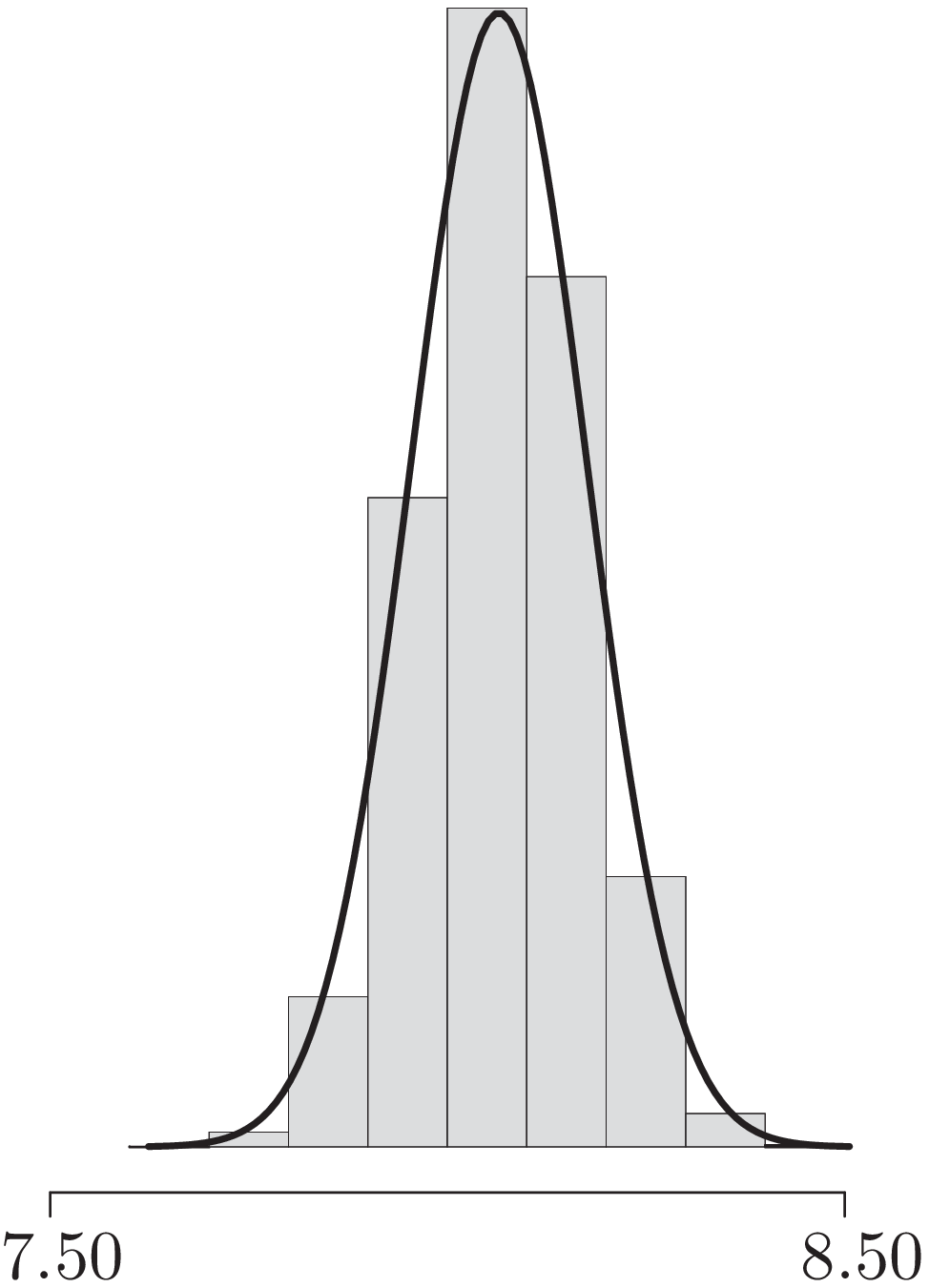}
    \caption{$c_{1111}$ of tensor~(\ref{eq:ca})}
\label{fig:c1111a}
\end{minipage}\hfill
  \begin{minipage}[b]{0.3\textwidth}
    \includegraphics[width=4cm,scale=0.5]{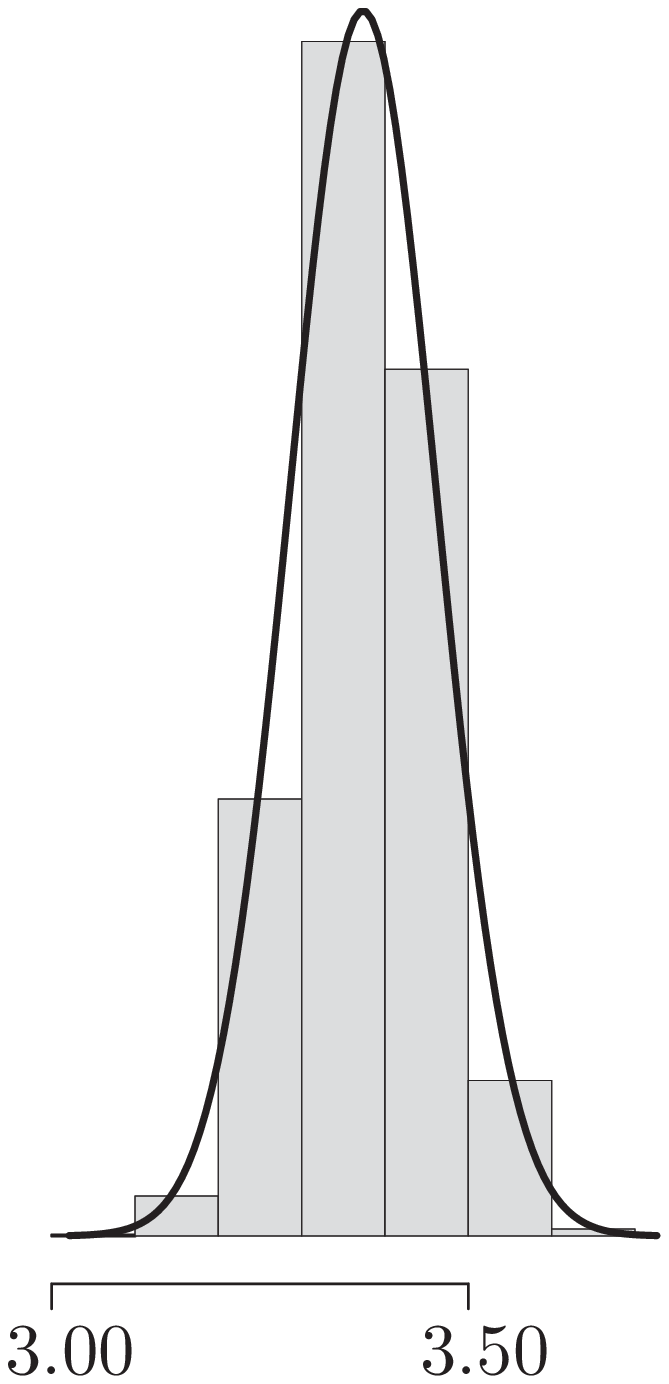}
    \caption{$c_{1122}$ of tensor~(\ref{eq:ca})}
\label{fig:c1122a}
\end{minipage}\hfill
\begin{minipage}[b]{0.3\textwidth}
    \includegraphics[width=6cm]{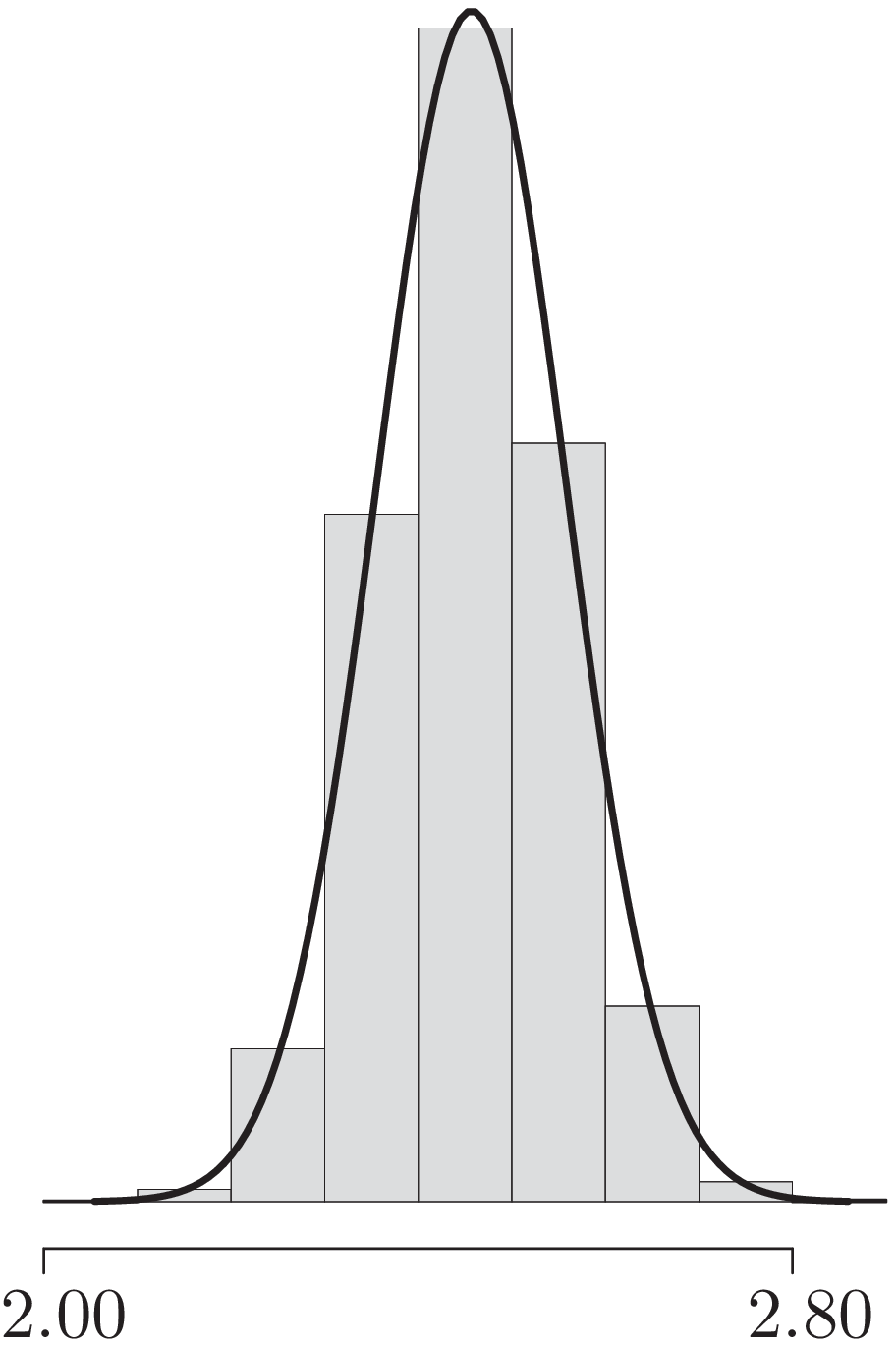}
    \caption{$c_{1133}$ of tensor~(\ref{eq:ca})}
\label{fig:c1133a}
\end{minipage}
\end{figure}

\begin{figure}[!htb]
\centering
  \begin{minipage}[b]{0.4\textwidth}
    \includegraphics[width=7cm]{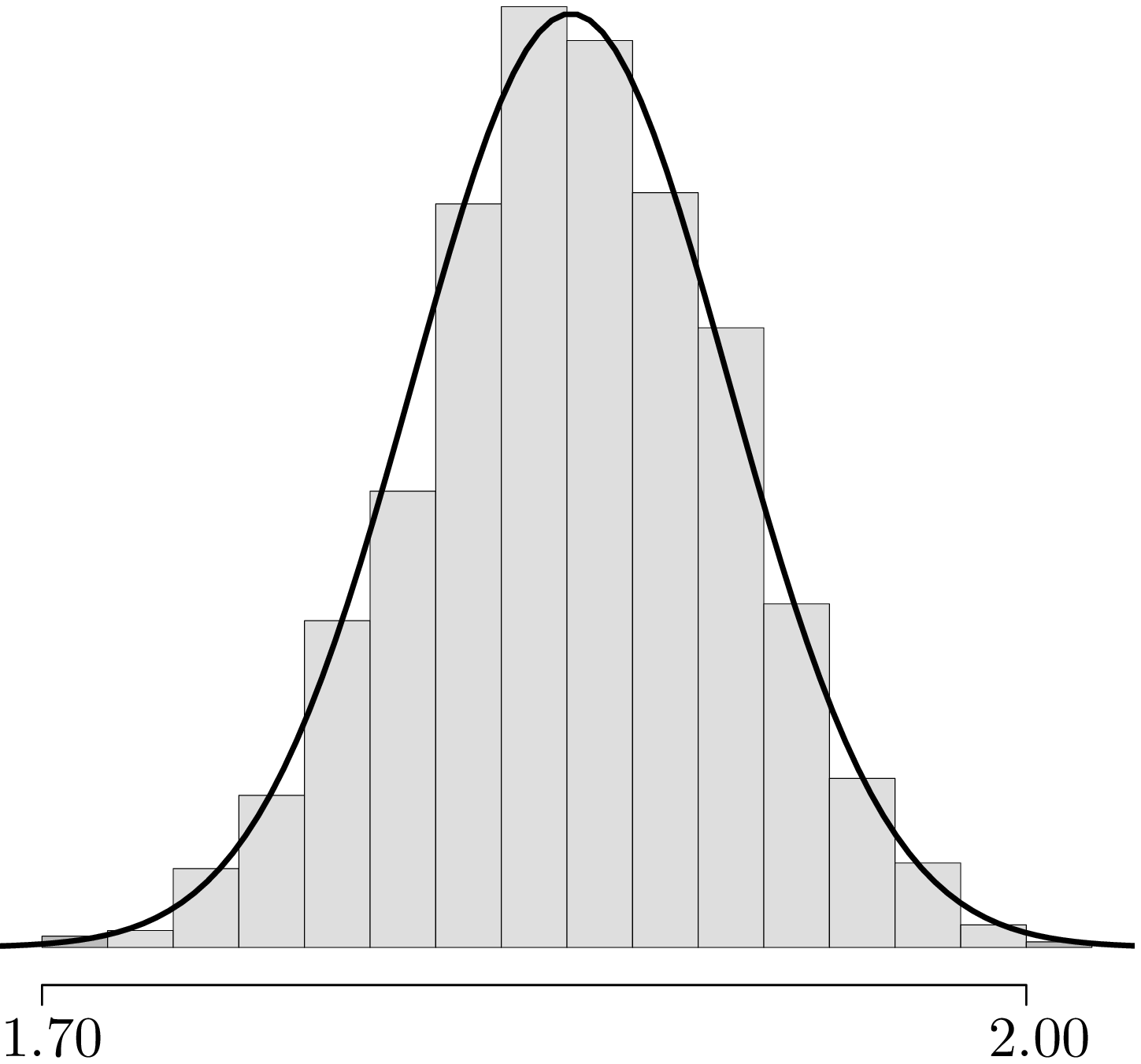}
    \caption{$c_{2323}$ of tensor~(\ref{eq:ca})\\\quad}
\label{fig:c2323a}
\end{minipage}\hspace{3cm}
  \begin{minipage}[b]{0.20\textwidth}
    \includegraphics[width=3.7cm]{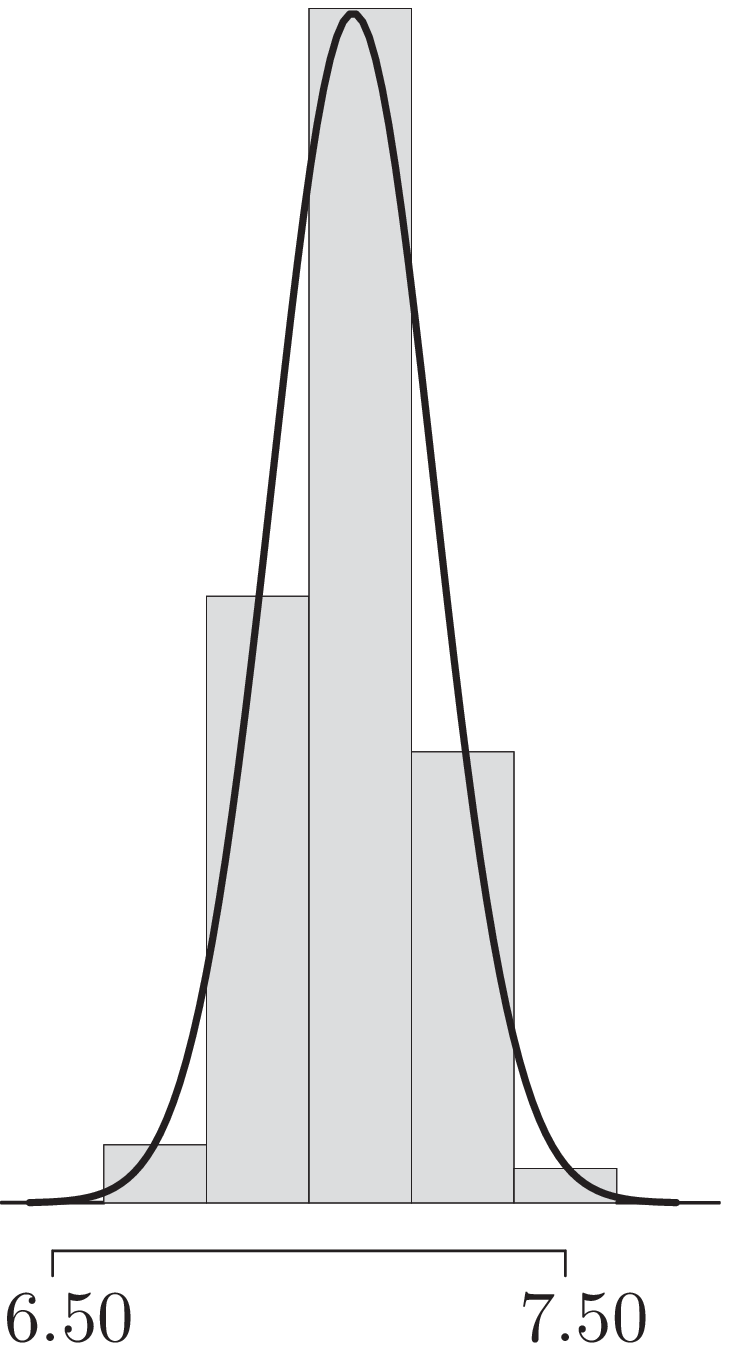}
\caption{$c_{3333}$ of tensor~(\ref{eq:ca})\\\quad}
\label{fig:c3333a}
\end{minipage}
\end{figure}

Components of an anisotropic tensor obtained from experimental measurements exhibit uncertainties due to measurement errors.
These uncertainties propagate to its symmetric counterparts.
The standard deviations of components of tensor~(\ref{Aniso}) are (Grechka, pers. comm., 2007)
\begin{equation}
\label{eq:standev}
\pm
\left[\begin{matrix}
0.1656 & 0.1122 & 0.1216 & 0.1176 & 0.0774 & 0.0741\\
0.1122 & 0.1862 & 0.1551 & 0.0797 & 0.1137 & 0.0832\\
0.1216 & 0.1551 & 0.1439 & 0.0856 & 0.0662 & 0.1010\\
0.1176 & 0.0797 & 0.0856 & 0.0714 & 0.0496 & 0.0542\\
0.0774 & 0.1137 & 0.0662 & 0.0496 & 0.0626 & 0.0621\\
0.0741 & 0.0832 & 0.1010 & 0.0542 & 0.0621 & 0.0802\\
\end{matrix}\right].
\end{equation}
These values do not constitute components of a tensor.
They are valid only in the coordinate system of measurements; rotation is not allowed.
Hence, to consider error propagation from tensor~(\ref{Aniso}) to tensor~(\ref{eq:ca}), there is a need for a simulation.
Probability distributions about the mean values of the components of tensor~(\ref{eq:ca})---obtained by a Monte-Carlo simulation (Danek et al. 2013)---are shown in Figures~\ref{fig:c1111a}, \ref{fig:c1122a}, \ref{fig:c1133a}, \ref{fig:c2323a}, \ref{fig:c3333a}.
Different histograms have different horizontal scales.

\begin{figure}[!htb]
  \begin{minipage}{0.4\textwidth}
    \includegraphics[width=\linewidth]{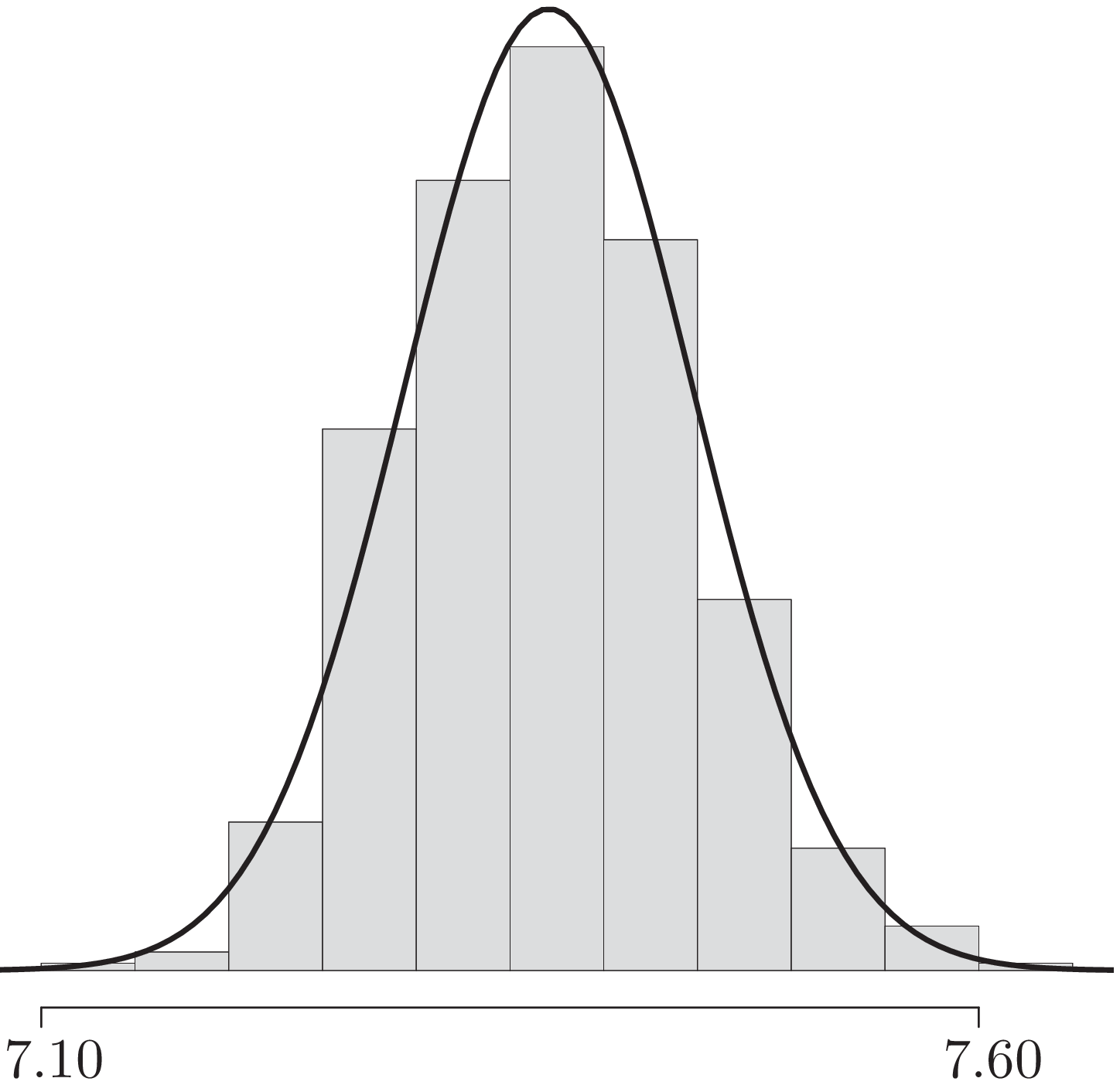}
    \caption{$c_{1111}$ of tensor~(\ref{eq:caIso36})}
\label{fig:c1111F36}
\end{minipage}\hfill
  \begin{minipage}{0.4\textwidth}
    \includegraphics[width=\linewidth]{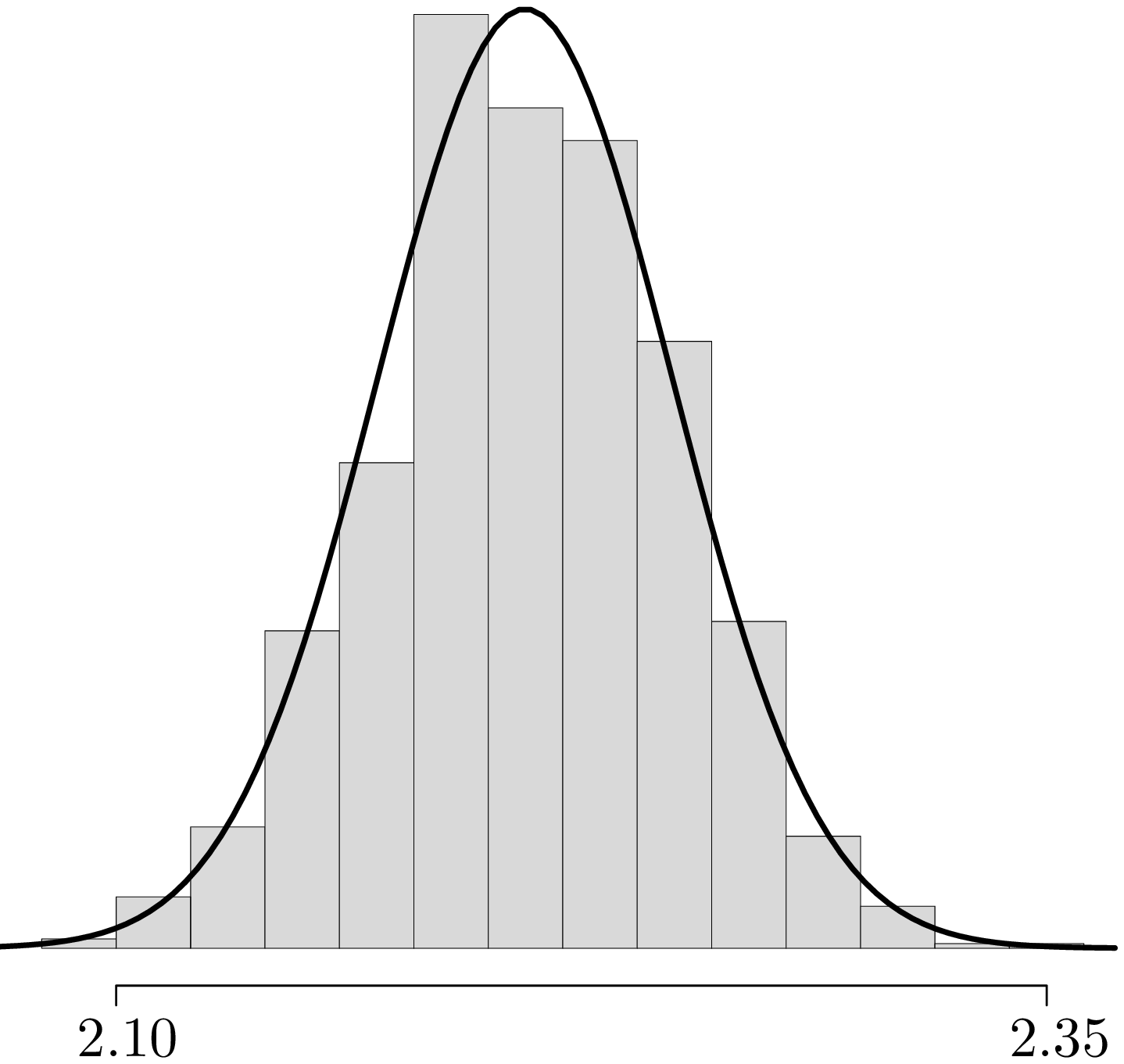}
    \caption{$c_{2323}$ of tensor~(\ref{eq:caIso36})}
\label{fig:c2323F36}
\end{minipage}
\end{figure}

\begin{figure}[!htb]
  \begin{minipage}{0.4\textwidth}
    \includegraphics[width=\linewidth]{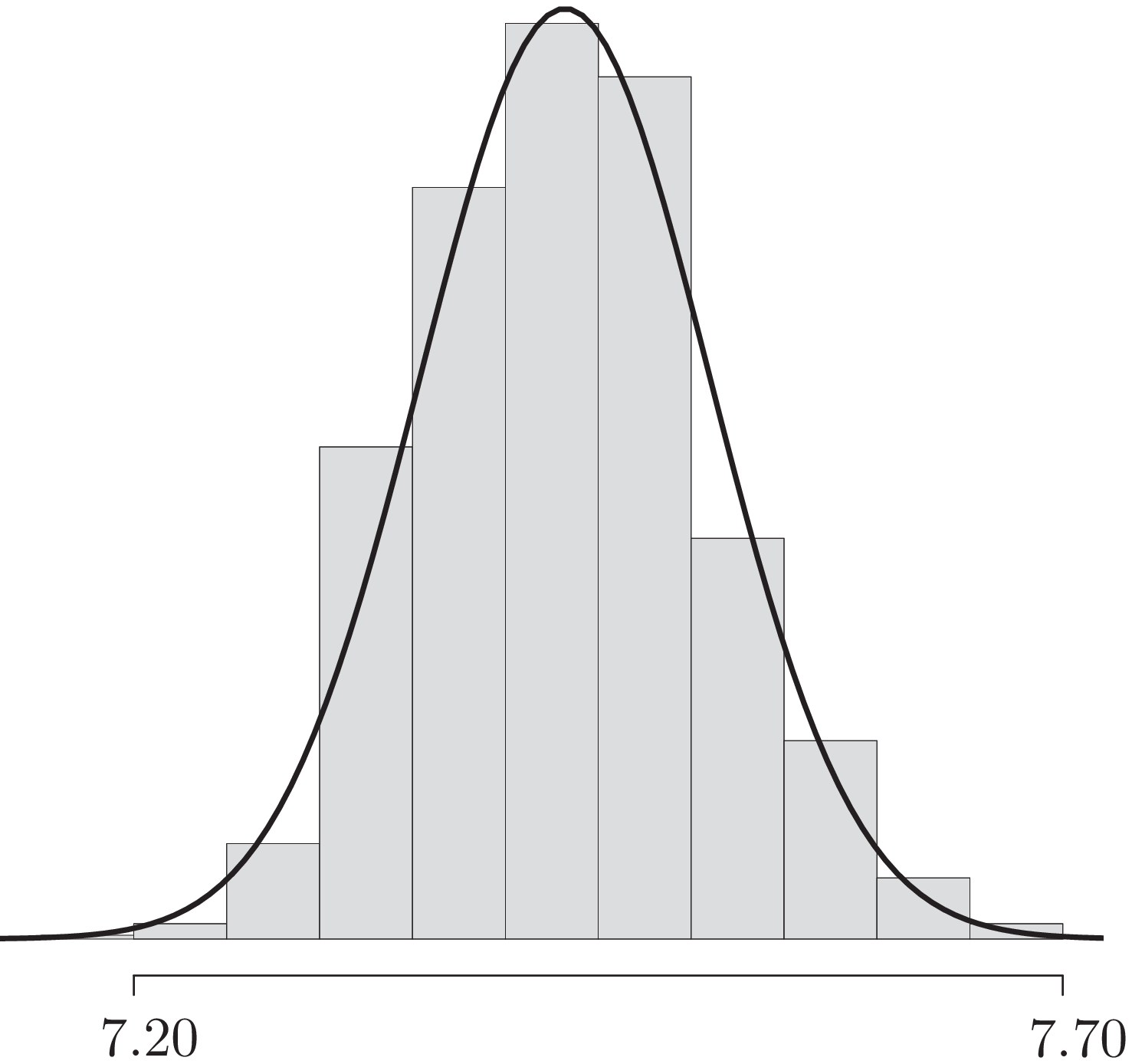}
    \caption{$c_{1111}$ of tensor~(\ref{eq:caIso21})}
\label{fig:c1111F21}
\end{minipage}\hfill
  \begin{minipage}{0.4\textwidth}
    \includegraphics[width=\linewidth]{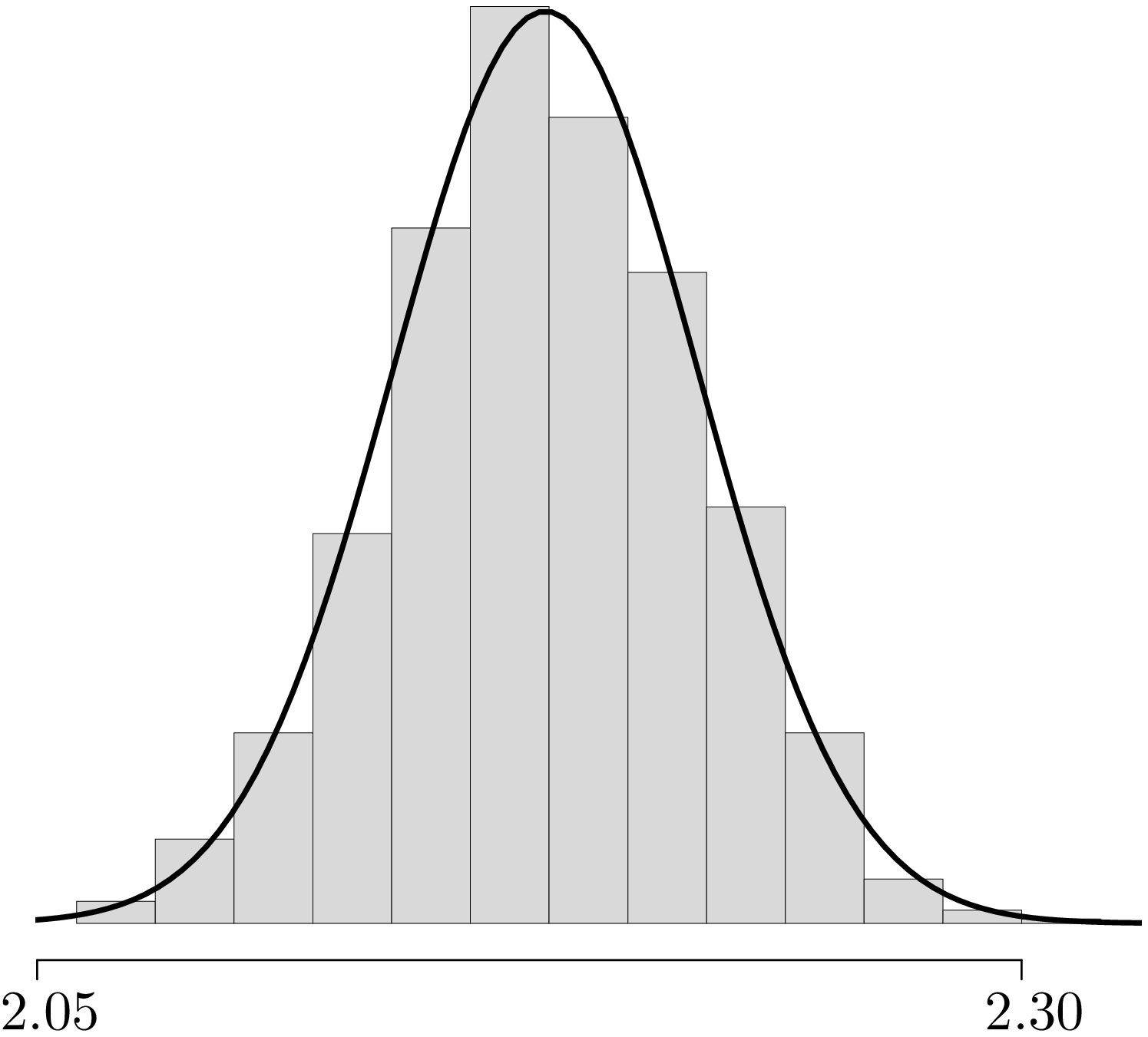}
    \caption{$c_{2323}$ of tensor~(\ref{eq:caIso21})}
\label{fig:c2323F21}
\end{minipage}
\end{figure}

\begin{figure}[!htb]
  \centering
  \begin{minipage}{0.4\textwidth}
    \includegraphics[width=\linewidth]{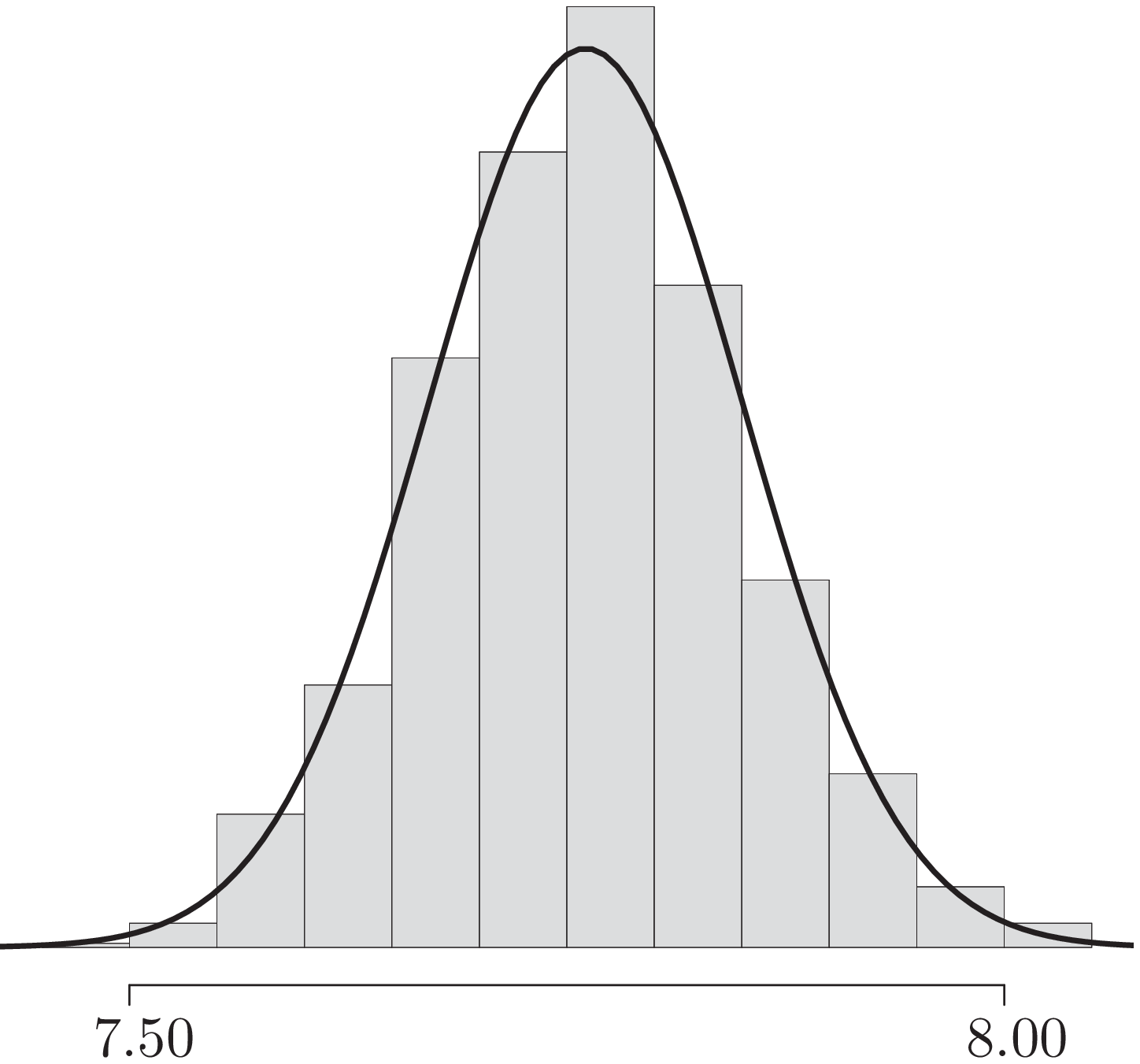}
    \caption{$c_{1111}$ of tensor~(\ref{eq:caIsoL2})}
\label{fig:c1111L2}
  \end{minipage}
  \hfill
  \begin{minipage}{0.4\textwidth}
    \includegraphics[width=\linewidth]{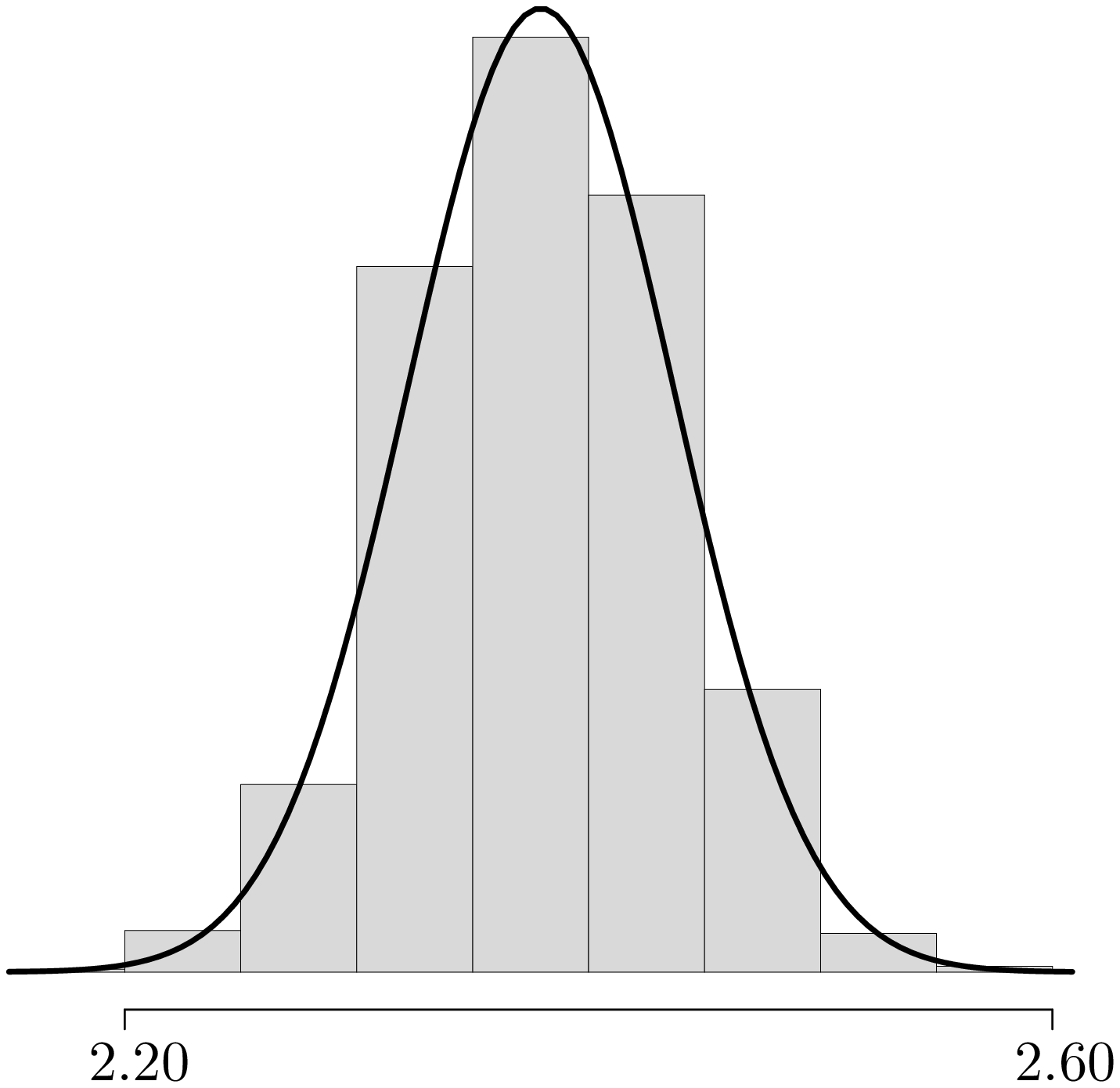}
    \caption{$c_{2323}$ of tensor~(\ref{eq:caIsoL2})}
\label{fig:c2323L2}
\end{minipage}
\end{figure}

The probability distributions of the two parameters for its isotropic $F_{36}$ counterpart are obtained in the same manner and shown in Figures~\ref{fig:c1111F36} and \ref{fig:c2323F36}. Their mean values are given in tensor~(\ref{eq:caIso36}).
The probability distributions of parameters for its $F_{21}$ counterpart are shown in Figures~\ref{fig:c1111F21} and \ref{fig:c2323F21}.
The probability distributions of parameters for its $\lambda$ counterpart are shown in Figures~\ref{fig:c1111L2} and \ref{fig:c2323L2}.



\section{Discussions and conclusions}

In Section~\ref{norms}, we consider several types of norms for obtaining---for a transversely anisotropic tensor---its closest isotropic counterpart.
We examine the Frobenius norms and the operator norm.
In Section~\ref{misfit}, we consider the slowness-curve $L_2$ fit to obtain such a counterpart.

The closeness to isotropy is norm-dependent.
Yet, the order of several tensors according to their closeness to isotropy for a particular norm remains the same for all norms.
However, as shown in Sections~\ref{sub:TensorC}, \ref{sub:Comparison} and \ref{sub:Thomsen}, given tensor~(\ref{eq:ca}) we can find another tensor---representative of common geological materials---such that one of them is closer to isotropy according to one norm and the other closer to isotropy according to another norm.

In view of Section~\ref{sub:Error}, we conclude that the results of the three norms and the slowness-curve fit are so similar to each other that their corresponding values might be indistinguishable in the context of measurement errors.
Thus, the choice of the closeness criterion might be of secondary importance.
Pragmatically, we might choose a Frobenius norm, since it offers analytical formul{\ae} to obtain an isotropic counterpart.
Both Frobenius norms result in similar counterparts, since they differ only by a weight doubling of the off-diagonal components, whose values are small.
Also, in view of this similarity, the preference in closeness to isotropy for the pairs of tensors discussed in Section~\ref{sub:Comparison} might be indistinguishable.

Performing a simple error-propagation analysis, we observe that---for Frobenius norms---probability distributions of the corresponding parameters are very similar to one another.
For the operator norm, however, the $c_{2323}$ distributions differ significantly.
This result might be a consequence of the properties of the operator norm, where only the largest among six eigenvalues is taken into consideration.

\section*{Acknowledgments}
We wish to acknowledge discussions with David Dalton, Michael G. Rochester and Theodore Stanoev, as well as graphical support by Elena Patarini.
This research was performed in the context of The Geomechanics Project
supported by Husky Energy. Also, this research was partially supported by the Natural Sciences and Engineering Research Council of Canada, grant 238416-2013, and by the Polish National Science Center under contract No.\ DEC-2013/11/B/ST10/0472.

\newpage
\section*{References}
\frenchspacing
\newcommand{\hd}{\par\noindent\hangindent=0.4in\hangafter=1}
\hd
Auld, B.A., {\it Acoustic Fields and Waves in Solids, Vol. 1\/}, Florida: Kreiger Publishing, 1978.
\hd
Bos, L., Slawinski, M.A., 2-norm effective isotropic Hookean solids,
{\it  J.~Elasticity}, {\bf 120}, 1, 1--22, 2014.
\hd
Danek, T., Kochetov, M., Slawinski, M.A., Effective elasticity tensors in the context of random errors, {\it J.~Elasticity}, {\bf 121}, 1, 4, 2015.
\hd
Danek, T., Kochetov, M., Slawinski, M.A., Uncertainty analysis of effective elasticity tensors using quaternion-based global optimization and monte-carlo method, {\it Q J Mech Appl Math}, {\bf 66}, 2, 2013.
\hd
Slawinski, M.A., {\it Wavefronts and rays in seismology: Answers to unasked questions\/},
World Scientific, 2016.
\hd
Thomsen, L., Weak elastic anisotropy, {\it Geophysics}, {\bf 51}, 10, 1954-1966, 1986.
\hd
Voigt, W., {\it Lehrbuch der Kristallphysics, Teubner}, Leipzig, 1910.

\end{document}